\documentclass[sigconf]{acmart}
\setcopyright{rightsretained}

\acmConference[NSPW'22]{New Security Paradigms Workshop}{October 24--27, 2022}{New Hampshire, USA}

\usepackage{balance}
\usepackage{amsmath}
\usepackage{graphicx}
\usepackage{listings}
\usepackage{xspace}
\usepackage{url}
\usepackage{siunitx}
\usepackage{booktabs} 
\usepackage{multirow}
\usepackage{dcolumn}
\usepackage{diagbox}

\usepackage{setspace}

\usepackage{caption}
\DeclareCaptionFont{white}{\color{white}}
\DeclareCaptionFormat{listing}{\colorbox[cmyk]{0.43, 0.35, 0.35,0.01}{\parbox{\columnwidth}{\hspace{15pt}#1#2#3}}}
\newlength{\thinline}
\newlength{\thickline}
\newcolumntype{.}{D{.}{.}{-1}}

\usepackage{subcaption}

\usepackage{microtype}
\usepackage{paralist}

\usepackage{enumitem} 

\newcommand{\iv}{I--\kern-.08emV\xspace}

\usepackage{color}
\usepackage{colortbl}
\usepackage{pifont}
\definecolor{darkgreen}{rgb}{0.0, 0.5, 0.0}
\definecolor{babyblueeyes}{rgb}{0.63, 0.79, 0.95}

\usepackage{multirow}

\definecolor{revcol}{rgb}{ 0 0 0}

\definecolor{crcol}{rgb}{ 0 0 0}

\newcommand{\camready}[1]{\textcolor{crcol}{#1}}
\newcommand{\rev}[1]{\textcolor{revcol}{#1}}

\begin{document}

\title[]{Side Auth: Synthesizing Virtual Sensors for Authentication}

\author{Yan Long}
\affiliation{%
  \institution{University of Michigan}
}
\email{yanlong@umich.edu}

\author{Kevin Fu}
\affiliation{%
  \institution{Northeastern University}
}
\email{k.fu@northeastern.edu}



\begin{abstract}
While the embedded security research community aims to protect systems by reducing analog sensor side channels, our work argues that sensor side channels can be beneficial to defenders.  This work introduces the general problem of synthesizing virtual sensors from existing circuits to authenticate physical sensors' measurands. We investigate how to apply this approach \camready{
and present a preliminary analytical framework and definitions for sensors side channels.} To illustrate the general concept, we provide a proof-of-concept case study  to synthesize a virtual inertial measurement unit from a camera motion side channel. Our work also provides an example of applying this technique to protect facial recognition against silicon mask spoofing attacks. Finally, we discuss downstream problems of how to ensure that side channels benefit the defender, but not the adversary, during authentication. 

\end{abstract}





\maketitle

\section{Introduction}

Sensor side channels enable an adversary to violate integrity of sensor outputs by influencing or controlling the sensor with transduction attacks~\cite{yan2020sok,giechaskiel2019taxonomy}, or to eavesdrop on sensitive information and compromise confidentiality  by exploiting flaws in  sensor and system designs~\cite{michalevsky2014gyrophone, bolton2021touchtone,anand2021spearphone, simon2013pin}. \camready{For example, the eavesdropping example PIN Skimmer \cite{simon2013pin} shows that adversaries can 
infer smartphone touchscreen inputs by exploiting side channel motion information captured by smartphone cameras.} While the security research community invested significant effort identifying and mitigating analog sensor side channels, our work argues that it can be {\bf beneficial to embrace, understand, and control analog sensor side channels instead of simply eliminating them}. 
\camready{This is motivated by our observation that such side channel information may also be used for authentication. For example, extensive research has been conducted on using dedicated motion sensors to capture smartphone touch dynamics for continuous  implicit user authentication~\cite{teh2016survey}. Relating it to PIN Skimmer, a natural question arises as to whether cameras support such authentication when dedicated motion sensors are not available.}  We thus propose and investigate the problem of how to utilize sensor side channels for defensive purposes such as  multimodal authentication by synthesizing virtual sensors from them.

\begin{figure}[!t]
	\centering
	\includegraphics[width=.45\textwidth]{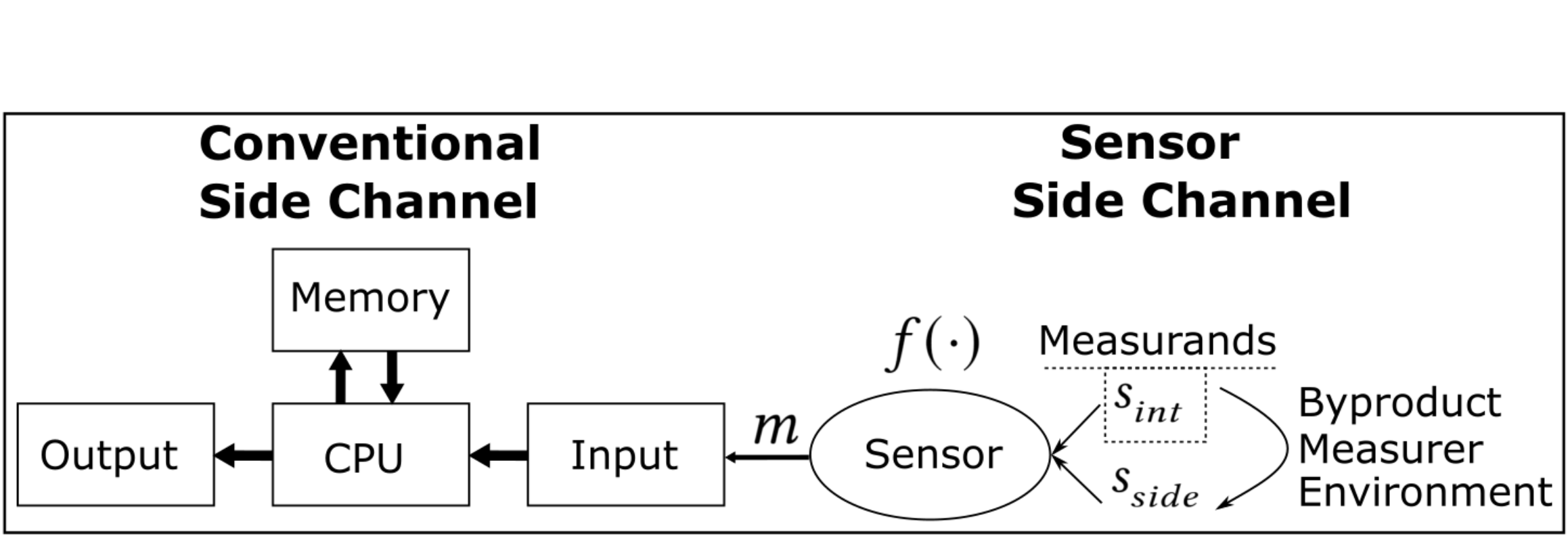}
	\vspace{-.1in}
	\caption{Sensor side channels are different from conventional side channels as they measure the measurement processes instead of computation processes. Sensor side channels can measure the byproduct, measurer, and environment to verify authenticity of intended sensor measurands.} 
	\label{figs:arch}
	\vspace{-.2in}
\end{figure}

Side channels are inherent to analog sensors' physics. There exist a considerable number of potential sensor side channels besides those revealed by transduction and eavesdropping attacks. However, most of these side channels are deliberately ``closed'' in the design phase by employing mitigation mechanisms such as calibration and noise reduction. It is foreseeable that sensor and system designers will also try to mitigate  newly discovered side channels. This work argues a different perspective and approach to embrace such sensor side channels.  If these side channels can be used in a beneficial way, we envision future designs allowing mitigation mechanisms to be strategically disabled or downgraded when needed such as during authentication sessions.

We provide a preliminary analytical framework for modeling analog sensor side channels and explaining the origins and characteristics of them. The framework categorizes sensor side channels according to their separability from intended signals and whether they have controllable mitigation mechanisms. Based on the framework, we define the problem of virtual sensor synthesis for multimodal measurand authentication and summarize three possible ways of applying this approach (Figure~\ref{figs:arch}). First, by verifying signatures of signal byproducts and asking the question ``What is the probability that Alice generated both the measurands and byproducts?'' Second, by verifying the person performing the measurement and asking  ``What is the probability that Alice generated the measurands if Bob was the measurer?'' Third, by verifying the environment of the measurement process and asking ``What is the probability that Alice generated the measurands if the measurement was taken in location B?''

A proof-of-concept case study further concretizes the concepts and related considerations by studying a camera motion side channel that enables cameras to sense out-of-sight motion. This side channel is caused by mechanical connections between camera devices and adjacent  objects in motion such as a hand holding the camera.  We propose a methodology for synthesizing virtual inertial measurement units (IMUs) from this side channel that can extract both inter-frame low-frequency  and intra-frame high-frequency motion information. The case study discusses this side channel's potential application in helping facial recognition systems defend against 3D silicon mask spoofing attacks by verifying postural hand tremor motion of the person holding the camera device. Preliminary test with 4 people suggests the camera motion side channel help reduce false positive rates by up to 87.5\%.  It also shows that disabling video stabilization  enables higher performance, emphasizing the benefits of strategically
disabling side channel mitigation mechanisms. Finally, we discuss the possible issues of temporarily opening sensor side channels during authentication and the directions future works may take to address the issues.  Our main contributions are summarized as follows:

\begin{itemize}
    \item A new paradigm to embrace and harness analog sensor side channels for defensive purposes via active control or influence of sensor side channels. 
    \item An analytical framework to enable definition and characterization of sensor side channels. The framework introduces the concept of virtual sensor synthesis for multimodal sensor measurand authentication. 
    \item A case study and methodology of synthesizing virtual IMUs from camera motion side channels for enhancing performance of facial recognition systems. Virtual IMUs decrease false positive rates when facial recognition is subjected to emulated 3D silicon mask spoofing attacks. 
\end{itemize}

\section{Background \& Related Work}

\camready{Side channels are unintended information output channels. The idea of synthesizing virtual sensors for authentication can be traced back to the general concept of using side channel information to identify people in forensic analysis. For example, forensic handwriting recognition allows one to determine the identity of a letter writer even  though the letter was never intended  to convey identity information. In the field of digital forensics~\cite{garfinkel2010digital}, the authenticity of digital evidence can be verified by cross-correlating file data and metadata with other contextual information. For example, fraudulent documents were reported to be detected by analyzing their choices of font~\cite{msfont}. Our paper investigates how we can design mechanisms to actively utilize side channel information in sensor readings for defending computer systems. We introduce research works related to this idea in this section. }

\subsection{Sensor Side Channel Based  Attacks }\label{sec:bg_attack}

Sensor side channels have been actively exploited in two lines of research, namely transduction  and eavesdropping attacks. This section provides some background and examples of these attacks. \rev{Different from these works that try to compromise information integrity and confidentiality of sensor systems, our work investigates how designers may defend sensor systems by actively controlling and utilizing sensor side channels.}

\textbf{Transduction Attacks.} Transduction attacks injects analog signals into sensors where victim sensor circuitry transduces an attacker’s malicious physical signals to untrustworthy sensor measurements. Such malicious physical signals can often be in different physical modalities (e.g., acoustic vs. optical) or frequency ranges (e.g., audible vs. ultrasound) than what the sensors are designed to sense. For example, Light Commands \cite{sugawara2020light} uses lasers to inject false speech signals into microphones. Works such as Walnut \cite{son2015rocking,trippel2017walnut} use acoustic injections to influence and control the output of MEMS gyroscopes and accelerometers.  Ghost Talk \cite{kune2013ghost} uses radio waves to inject audio signals into microphones. An SoK and a survey~\cite{yan2020sok,giechaskiel2019taxonomy} provide  a comprehensive review of theses attacks and corresponding mitigation mechanisms.

\textbf{Eavesdropping Attacks.} Sensor side channels are also exploited for eavesdropping out-of-band information. For example, PIN Skimmer \cite{simon2013pin} used a camera-based side channel to infer PIN inputs by exploiting correlations between smartphone camera orientations and tapping locations. Several works such as Gyrophone~\cite{michalevsky2014gyrophone, bolton2021touchtone,anand2021spearphone} use smartphone IMUs which contain accelerometers and gyroscopes to eavesdrop speech by exploiting side channels enabled by signal aliasing in analog-to-digital converters (ADCs).

\subsection{\rev{Using Conventional Non-sensor} Side Channels For Defensive Purposes}
\rev{Although we could not find related research that investigates the concept of utilizing sensor side channels for defending a system, we found that} a few previous works explored using  \textbf{non-sensor} side channels for machine-to-machine authentication. \cite{sakiyama2016physical} proposes to use the key-dependent side channel information in wireless communication channels to enhance existing cryptographic protocols. \cite{dabosville2019bright} presents an extension by analyzing practical and security issues of the protocol in \cite{sakiyama2016physical}
and providing fixes. Compared to them, this work focuses on the  concept of  sensor side channel and authentication of sensor's measurands instead of computation-generated information as the previous works did. \rev{Non-sensor} side channels are also utilized in other applications such as code-execution monitoring and intrusion detection~\cite{park2019leveraging,aubel2017side,clark2013wattsupdoc}.







\section{Problem Formulation}\label{sec:prob}

This section defines the problem of using sensor side channels for measurand authentication. \rev{Our paper proposes the concept of using sensor side channels for authentication as a new direction of research for the community. We also fill a gap by suggesting a mathematical definition of sensor side channels, beginning} with a framework for defining and categorizing sensor side channels. We then introduce the problem of synthesizing virtual sensors and using them for multimodal sensor measurand authentication.

\subsection{Sensor Side Channel Analytical Framework}

\subsubsection{Sensor}\label{sec:side_sensors} A sensor, or transducer, can be modeled as a function that maps physical measurands  to digital measurements over time. A measurand is a quantity that a sensor intends to
measure~\cite{goepel1994sensors}.  Different types of sensors are designed to measure different modalities of measurands such as sound, temperature, motion, etc. Users who are informed of the apparent purpose and specifications of sensors often see a sensor as the following function over a single variable of the measurand:
\begin{equation} \label{eq:sensor_1}
    m = f(s_{_{int}})
\end{equation}
where $m$  and $s_{_{int}}$ denote the digital measurements and analog measurand respectively and $f(\cdot)$ denotes the sensor. 

\subsubsection{Sensor Side Channels}\label{sec:side_causes} Although Equation~\ref{eq:sensor_1} provides average sensor users a clean and easy abstraction, actual sensor implementations are much ``dirtier' and introduce numerous hidden variables to the equation that result in unintended components in measurement $m$. For instance, every conductor wire can be regarded as an unintentional antenna, leading to side channels that convert electromagnetic energy to measurements of non-electromagnetic sensors \cite{kune2013ghost, tu2019trick}. In this case, hidden variables related to electromagnetic energy in the environment should be added to Equation~\ref{eq:sensor_1}.  Another example of such variables is  temperature. Semiconductors made of silicon are inherently sensitive to heat due to its  ability to excite electrons. So technically, Equation~\ref{eq:sensor_1} should also include temperature as a variable. Electromagnetic energy and temperature are just examples of hidden variables associated to the underlying physical characteristics of devices. There are also hidden variables caused by design flaws and uncontrollable variations in the manufacture processes. Thus, Equation~\ref{eq:sensor_1} should be modified to enable a side channel-aware modeling of sensors:

\begin{equation} \label{eq:sensor_2}
    m = f(s_{_{int}}, s_{_{side}}), \quad s_{_{side}} = [s_{v_1},s_{v_2}, ...]
\end{equation}
where $s_{_{side}}$ represent the set  of all these hidden variables that can potentially lead to side channels attacks.

The comparison between Equation~\ref{eq:sensor_1} and \ref{eq:sensor_2} shows that the gap between users' understanding and sensors' actual  implementation gives birth to sensor side channels. \camready{On a high level, we believe the gap can also be attributed to the insufficient specifications of legitimate and illegitimate sensor behaviors in the existing system's security policies. Note that this differs from conventional non-sensor side channels where side channels bypass the clearly specified security policies~\cite{gligor1994guide}: there are often no dedicated security policies for sensors yet in existing sytems.   Building upon previous side channel research~\cite{zhou2005side, spreitzer2017systematic},  we tentatively define sensor side channels as the following}: 
\begin{itemize}
    \item \camready{\textit{A sensor side channel is a communication channel that allows someone to recover secret information using  unintended sensor measurement components  in a way that violates the associated system's security expectations.}} 
\end{itemize}

Sensor side channels are sometimes more conceptually difficult to recognize than conventional non-sensor side channels such as differential power analysis channels. The reason is that non-sensor side channels are used to mainly measure computation processes where there exists a clear boundary between  computation and measurement whereas  sensor side channels are used to measure the measurement processes themselves (Figure \ref{figs:arch}).

A possible way of identifying sensor side channels is to test the hypothesis that the analog signal of a variable $v_i$ correlates with $m$ with certain significance, i.e., 
\begin{equation} \label{eq:def}
    |Corr(m, s_{v_i})| > \alpha, \quad s_{v_i} \in s_{_{side}}
\end{equation}
where $\alpha$ is a threshold value. Note that this work does not discuss the actual choice of threshold values and correlation functions since they can be flexible depending on the actual application scenarios and security requirements.  In cases where it is challenging to project $m$ and $s_{v_i}$ to the same vector space in order to  compute correlation scores, other methods such as supervised classification can also be used if $s_{v_i}$ can be converted into data labels.

\subsubsection{Separability and Controllability}\label{sec:side_controllability}  The unintended components in the measurements are caused by the existence of $s_{_{side}}$ and can be either separable or inseparable from the intended components. The separability between the intended and unintended components is the key that decides whether a side channel can be mitigated and controlled or not. Conceptually, separable components can be defined as the following: there exists at least one function $\Tilde{f}(\cdot)$ that can break $m$  down into intended and unintended  components such that those components only correlate (with significance) with the measurand and other hidden variables respectively, i.e., 
\begin{align}\label{eq:sep}
    \exists \Tilde{f}(\cdot) \quad &s.t. \quad \Tilde{f}(m) = [m_{_{int}}, m_{_{side}}], \quad m_{_{side}} = [m_{v_1},m_{v_2}, ...], \nonumber\\ 
    &|Corr(m_{_{int}}, s_{_{int}})| > \alpha_{i1}, |Corr(m_{v_i}, s_{v_i})| > \alpha_{i2}, \nonumber\\
    &|Corr(m_{_{int}}, s_{v_i})| < \beta_{i1}, |Corr(m_{v_i}, s_{_{int}})| < \beta_{i2}
\end{align}

When a sensor side channel has separable components, we say it is a \textit{separable side channel}. Separability is decided by sensor implementation $f(\cdot)$. Side channels with inseparable components in existing sensor implementations led to the various unsolvable attacks against sensors (Section~\ref{sec:bg_attack}) because designers cannot extract only the intended components.

Theoretically, those with separable components can be mitigated by mechanisms referred to as compensation, calibration and noise reduction. Such mitigation mechanisms can be abstracted as another function $g(\cdot)$ that suppresses the unintended components in the output of $\Tilde{f}(\cdot)$, i.e., $g(\Tilde{f}(m)) = m_{_{int}}$. If the mitigation mechanisms can be both turned  on and off, the user of the sensor system then have full control of the sensor side channel. We call such a sensor side channel \textit{controllable}:
\begin{itemize}
    \item \textit{A controllable sensor side channel is one whose corresponding  unintended measurement component is separable from the intended component and can be suppressed by a mitigation mechanism that can be enabled and disabled.} 
\end{itemize}

\subsubsection{Examples}\label{sec:side_example} We provide some existing examples of each category of sensor side channels  to shed light on the differences and possible future evolution. 

\textbf{Inseparable.} The Gyrophone  eavesdropping attack~\cite{michalevsky2014gyrophone} and its follow-up works~\cite{bolton2021touchtone,anand2021spearphone} use an aliasing-enabled inseparable acoustic side channel in smartphone IMUs to recover speech. These IMUs have intended acceleration and angular velocity measurands mostly under the frequency range of human speech. However, due to the lack of effective analog low-pass filtering before the ADC, aliases of the high-frequency speech signals exist in the output of ADC and enable adversaries to recover speech information. Furthermore, the aliases cannot be separated from the intended motion signals since they are in the same frequency range. Intuitively, adding analog filters to the sensors make this acoustic side channel separable. In order to be controllable,  the sensor  API may further allow CPU to enable and disable the filters.  

\textbf{Separable But Uncontrollable.} Those seemingly intact sensors that have not been reported vulnerable to side channel-based attacks also have inherent side channels, but just in a suppressed manner thus these channels are not exposed to attackers. Take sensors' heat sensitivity mentioned in Section~\ref{sec:side_causes} as an example. MEMS humidity sensors, gyroscopes, accelerometers, etc., are widely equipped with temperature-compensated designs or online thermal calibration procedures~\cite{yang2018novel, garcia2020low, chen2008mems}. It can be anticipated that if the compensation and calibration mechanisms can be temporarily disabled, these sensors' measurements will exhibit significant correlation with the ambient temperature. In this way, the separable side channel becomes controllable.  

\textbf{Controllable.} There already exist sensors with controllable side channels. A good example is handheld cameras getting equipped with video stabilization mechanisms. Camera motion is often regarded as side effects that degrade the quality of the intended signal, i.e., the scene in the field of view of the camera~\cite{yang2009robust}. Video stabilization mechanisms, including electronic image stabilization (EIS) and optical image stabilization (OIS), etc., are implemented to mitigate these side effects by optically or electronically reducing the unwanted image scene movements caused by camera motion. Many operating systems such as Android allow app developers to choose if these video stabilization mechanisms will be turned on or off \rev{when the underlying camera hardware offers the API to control it}.  \rev{However, it is worth noting that such existing controllable side channels are most likely byproducts of OS designers' conventions of  providing more fine-grained interfaces, especially for open-source OS like Android which allows users to control EIS and OIS separately. In contrast, iOS does not allow explicit and separate  control of EIS and OIS.  Such a large degree of control is provided to support more potential use cases and enhance usability. For example, users may want to disable smartphone's built-in optical image stabilization when using an external gimbal because the two can interfere with each other and produce extra image distortions~\cite{iphoneOIS}. To the best of our knowledge, these existing controllable side channels have not been explored to enhance the security of systems.}

\subsubsection{Summary} It is possible to convert existing inseparable or uncontrollable side channels into controllable side channels by improving sensor designs, as has been suggested by the increasing popularity of video stabilization in cameras. Thus, it is important to think from a perspective of technology development when considering benefits of sensor side channels. Furthermore, protecting physical sensors from side channel attacks often already means transforming inseparable side channels to be separable. With some additional effort of making mitigation mechanisms controllable instead of forever-on, sensor side channels can be used in a beneficial and controlled manner.  The following discussions assume sensors have controllable side channels.

\subsection{Measurands Authentication Using Synthesized Virtual Sensors}

\subsubsection{Virtual Sensor Synthesis.} \label{sec:syn_syn}
A virtual sensor is a function that maps $m$ to  $m_{v_i}$. Ideally, the construction of $\Tilde{f}(\cdot)$ in Equation~\ref{eq:sep} already presents such an overarching function that can measure both the intended and side channel components. Such construction is apparently challenging since it needs to consider all possible side channels. Actual implementations can reduce the level of challenge by focusing on maximizing  $|Corr(m_{v_i}, s_{v_i})|$ and $-|Corr(m_{v_i}, s_{_{int}})|$ for only the set of targeted hidden variable $\{v_i\}$. We denote such a function specifically crafted for $\{v_i\}$ as $\Tilde{f}_{\{v_i\}}$ and call them virtual sensor functions.

\subsubsection{Problem Definition}\label{sec:syn_probdef}
We define the problem as a binary hypothesis test in a comparative manner by first referencing to the unimodal authentication on the physical sensor's measurand alone. Without virtual sensors, objects in Equation~\ref{eq:sensor_1} including $m$, $s_{_{int}}$, and $f$  are all that the designer of the authentication system can perceive. Let there be a measurand with a true identity $L$ and  a claimed identity  $\Tilde{L}$. The $H_1$ and $H_0$  hypotheses are $\Tilde{L} = L$ and  $\Tilde{L}  \neq L$ respectively. Denote the unimodal authentication system as $\mathcal{A}_u: m \rightarrow \{1, 0\}$, where it declares $H_1$ and $H_0$ when outputting 1 and 0 respectively. We can then define the total error of the unimodal system $E_u$ as 
\begin{align} \label{eq:error_u}
    E_u &= c_1\mathbb{P}[\text{declare } H_1 | H_0] + c_2\mathbb{P}[\text{declare } H_0 | H_1] \nonumber\\
    &= c_1\mathbb{E}[\mathcal{A}_u(m) | H_0] + c_2\mathbb{E}[1-\mathcal{A}_u(m) | H_1]
\end{align} 
where $\mathbb{P}[\cdot | \cdot]$ and $\mathbb{E}[\cdot | \cdot]$ denotes conditional probability and expectation respectively, $c_1$ and $c_2$ denote the cost coefficients for false positive and false negatives respectively.

Similarly, a multimodal authentication system with $n$ synthesized virtual sensors can be denoted as  $\mathcal{A}_m: [m_{_{int}}, m_{v_1}, ..., m_{v_n}] \rightarrow \{1, 0\}$. The total error $E_m$ is defined as 
\begin{align}\label{eq:error_m}
    E_m &= c_1\mathbb{E}[\mathcal{A}_m([m_{_{int}}, m_{v_1}, ..., m_{v_n}]) | H_0] \nonumber \\
    &+ c_2\mathbb{E}[1-\mathcal{A}_m([m_{_{int}}, m_{v_1}, ..., m_{v_n}]) | H_1]
\end{align}

As a result, the problem of synthesizing virtual sensors to authenticate the measurand in a multimodal manner can  be defined as:
\begin{itemize}
    \item \textit{Constructing  virtual sensor functions $\Tilde{f}_{\{v_i\}}$  and  multimodal authentication system $\mathcal{A}_m$  such that better performance  is achieved for measurand authentication, i.e., $E_m - E_u < 0$.}
\end{itemize}


\subsubsection{Security Properties} \label{sec:secprop}
Although  multimodal authentication using synthesized virtual sensors look similar to that using multiple physical sensors, it provides two different security properties. 

First, it works with existing devices and media that only have a single physical sensor's data. Although high-end devices like smartphones are equipped with multiple physical sensors, there still exist lower-end devices that only serve a single purpose such as ultrasonic proximity detectors and humidity monitors. Furthermore, sometimes it is needed to verify the identity of an object such as a photograph that has already been generated with only a single sensor. In this case, synthesized virtual sensors can extract additional  information in a retrospective way. 

Second, it potentially provide more robustness against spoofing attacks on individual sensors. The level of attack difficulty depends on the complexity of $\Tilde{f}$, i.e.,  how difficult it is to decouple and then modify different measurement components. Using multiple individual sensors such as cameras, accelerometers, etc., is equivalent to having a $\Tilde{f}$ that does not need to decouple anything at all since the inputs already separated.  Conceptually, if we regard the measurements corresponding to different virtual or physical sensors as random variables, we can then regard their variances and covariances as the entropy provided for authentication~\cite{mukher1986functional}. Virtual sensors potentially provides more entropy because the coupling between them adds to the covariances. Such entropy originates from the intrinsic physics of sensors. 

\subsubsection{Application} The general problem definition can be applied to different sources of side channel variables whose signatures correlate with the claimed identity of the measurands. Depending on the sources, we believe synthesized virtual sensors can be applied in the following three ways to verify authenticity of measurands. 

\textbf{Byproduct Verification.} A physical process generating intended measurands is likely to generate other forms of energy as  byproducts. Let us explore the example of a loudspeaker that replays a person Alice's speech recordings while a nearby microphone is listening to this replay. Say there is someone claiming the speech audio collected by the microphone is coming from Alice herself speaking live and an investigator tries to verify this claim.  The investigator finds out that the loudspeaker also generates unintended, secondary byproducts in the form of structure-borne vibrations, electromagnetic emission, heat, etc., which may be sensed by virtual sensors synthesized from the microphone's side channels. So, if these byproducts exist, the investigator knows it is not likely a legitimate recording of Alice's voice. In this case, the core authentication question can be summarized as \textit{``What is the probability that Alice generated both the measurands and byproducts?''}

\textbf{Measurer Verification.} A Measurer is the person who makes measurements with a physical sensor. Measurers themselves generate unintended emissions taking the form of  physical signals containing certain signatures that correlate with the identity of measurands. For example, say there exists an unmodified photo of a person who is claimed to be Alice and an investigator tries to verify this claim. The investigator managed to find out that the camera operator who took this photo, i.e., the measurer was Bob because Bob was speaking when he took the photo and his speech induced identifiable image blurs through a camera motion side channel.  If the investigator also knows that Bob has never been in the vicinity of Alice, then the investigator knows the person in the photo is not Alice. Obviously, measurand authentication through measurer verification may require higher-level contextual information compared to byproduct verification. The core authentication question is \textit{``What is the probability that Alice generated the measurands if Bob was the measurer?''}

\textbf{Environment Verification.} Similar to measurer verification, verifying the environment surrounding measurands also allows one to authenticate the measurands. Take the same example above. Say  the photo has a temperature side channel that shows the ambient temperature was 104°F/40°C at the time of generating the photo, pointing to a location B. If the investigator knows Alice has never been in location B, then the investigator knows the person in the photo is not Alice. The core authentication question  is \textit{``What is the probability that Alice generated the measurands if the measurement was taken in location B?''}



\section{Case Study} \label{sec:case_tremor}
The case study demonstrates how to use camera motion side channels (Section~\ref{sec:side_example}) to synthesize virtual IMUs that can collect postural hand tremor information for measurand authentication in facial recognition applications. It can be regarded an example of both byproduct and measurer verification. 

\subsection{Primer} \label{sec:case_primer}

\subsubsection{Postural Tremor Information.} Tremor is the involuntary rhythmic movement of a human body part caused by reciprocal innervations of muscles. Such involuntary movements are present in all people, with those found in healthy people and disease conditions (e.g., Parkinson disease) classified as physiological and pathological tremor respectively~\cite{timmer1996quantitative}. Clinical research finds that tremors measured by accelerometers can effectively predict the category of tremors. Some works further show that hand tremors measured by accelerometers and gyroscopes are unique to an individual and stable over time, suggesting the feasibility of using tremors as a biometric for personal identification~\cite{miu2016person,dun2019replicability}.

\subsubsection{Threat Model.}
We study a threat model of spoofing attack against smartphone facial recognition systems where imposters are assumed to launch a silicone face mask spoofing attack~\cite{ramachandra2019custom}. To  better show the effectiveness of the synthesized IMUs,  we further assume the silicone mask perfectly mimics the face of the victims. During the attack,  the imposter wears the silicone mask and holds the victim's smartphone for authentication. Our objective is to extract camera motion from videos \rev{that represents the postural hand tremor of users}  to defend against such perfect silicone mask attacks. 

\rev{It is worth noting this particular case study's threat model requires users to hold their phones in their hands during facial recognition as the contact between their phones and hands provides a propagation path for the vibration information of hand tremor. We believe this is also the most frequent situation seen in smartphone-based facial recognition applications. Nevertheless, there do exist some circumstances where users may want to place their phone on a table during authentication. Our tremor recognition with synthesized virtual IMUs will not work in this case due to the lack of camera motion. Similarly, a spoofing attacker cannot authenticate successfully in this case without providing the camera the correct motion. To enable users to authenticate without holding their phones, we believe future works may look into other sensor side channels that acquire a different type of user biometric information such as body-radiated electromagnetic/heat energy without requiring direct contact with the phone. }

\subsection{Synthesis Methodology}

Different methodologies can be used to synthesize virtual IMUs from camera motion side channels. For example, a completely model-based methodology requires understanding $f(\cdot)$ and $\Tilde{f}(\cdot)$. Although the most accurate, it requires  thorough understandings of every targeted camera system and is challenging. Another possible methodology is to completely rely on neural network to process the raw videos and let the network figure out $\Tilde{f}_{\{v_i\}}$, which is similar to previous work of inferring sounds from object motions in videos~\cite{owens2016visually}. This methodology requires intensive computation resources and data collection. This work focuses on the middle ground by investigating a model-informed methodology that constructs $\Tilde{f}_{\{v_i\}}$ based upon the concepts of image registration. Image registration is the process of overlaying two or more images of the same scene that are taken at different times, from different viewpoints, and/or by different sensors~\cite{zitova2003image}. The methodology aims to extract both inter-frame motions and intra-frame motions.

\begin{figure}[!t]
	\centering
	\includegraphics[width=.45\textwidth]{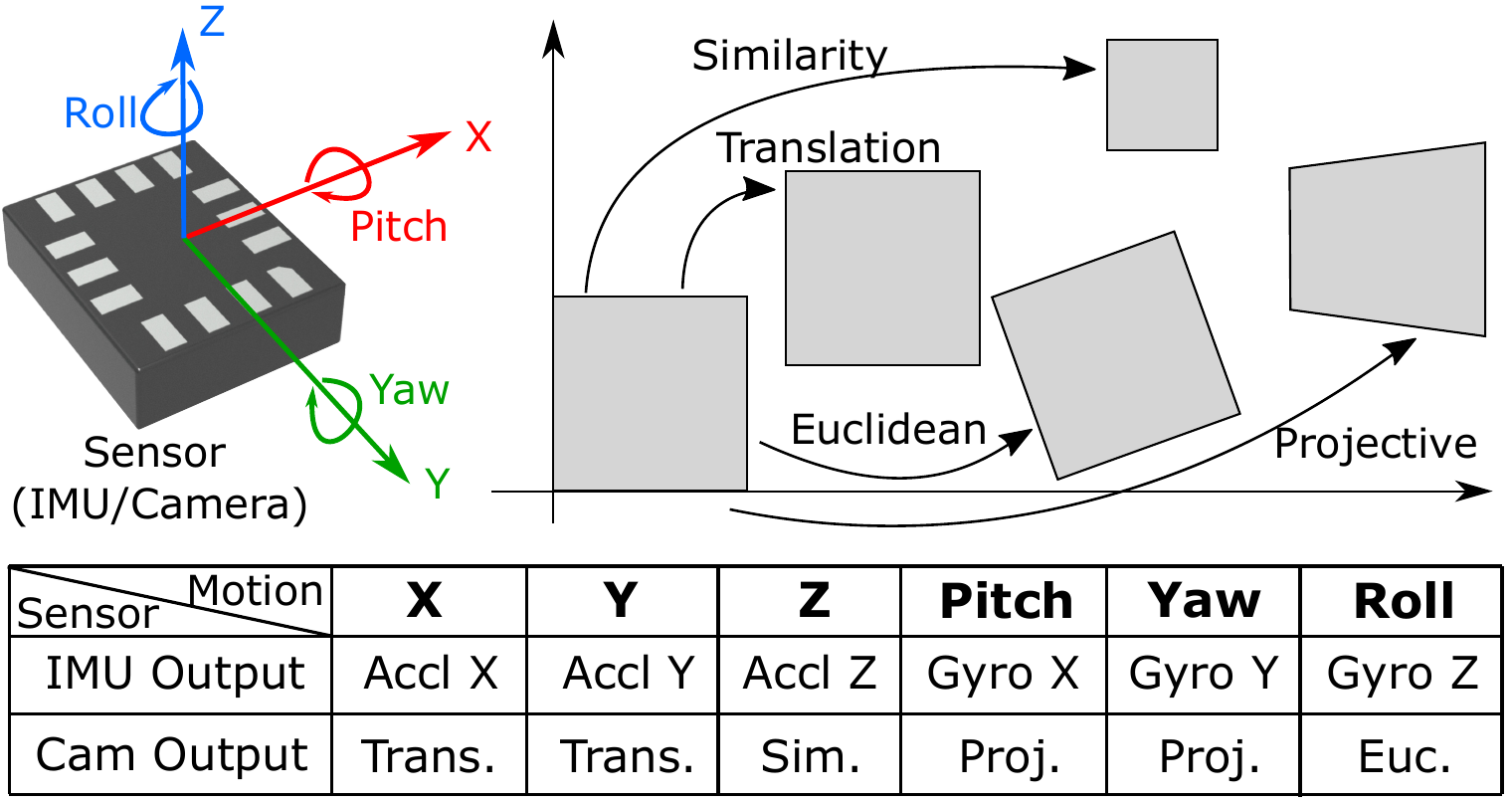}
	\vspace{-.1in}
	\caption{Types of 2D image transformations corresponding to the type of camera motion and  motion readings measured by physical IMUs. } 
	\vspace{-.2in}
	\label{fig:trans}
\end{figure}

\subsubsection{Understand Motion Modulation.} 
To construct $\Tilde{f}_{\{v_i\}}$, the first step is to understand how motion signals are modulated onto image streams. We analyze the motion modulation process from two different perspectives.

\textbf{Frame Transformation.} The frame transformation perspective considers changes of the frames subjected to camera motions as 2D image transformations. Figure~\ref{fig:trans} shows the possible image transformations corresponding to motion on each one of the six real-world axes and the measurements of physical IMUs. As a result, motions that can be measured by IMUs can also be mapped to inter-frame variations of the camera videos. 

\textbf{Rolling Shutter.} Besides inter-frame variations, the rolling shutter property of most cameras on portable devices can generate intra-frame variations that embed high-frequency motion. Rolling shutter is the shutter mechanism of commercial CMOS cameras, which exposes and samples the rows of an image sensor sequentially instead of simultaneously as in a global shutter~\cite{liang2008analysis}. If viewing the possible 2D image transformations as bases, rolling shutter combine multiple transformations into a single frame. It increases the effective sample rate of the motion signals provided by the camera side channel.

Based on the knowledge of how camera motion is modulated onto images, two corresponding categories of virtual IMU synthesis methods are introduced next to measure low-frequency and high-frequency information respectively.

\begin{figure*}[!t]
	\centering
	\includegraphics[width=.95\textwidth]{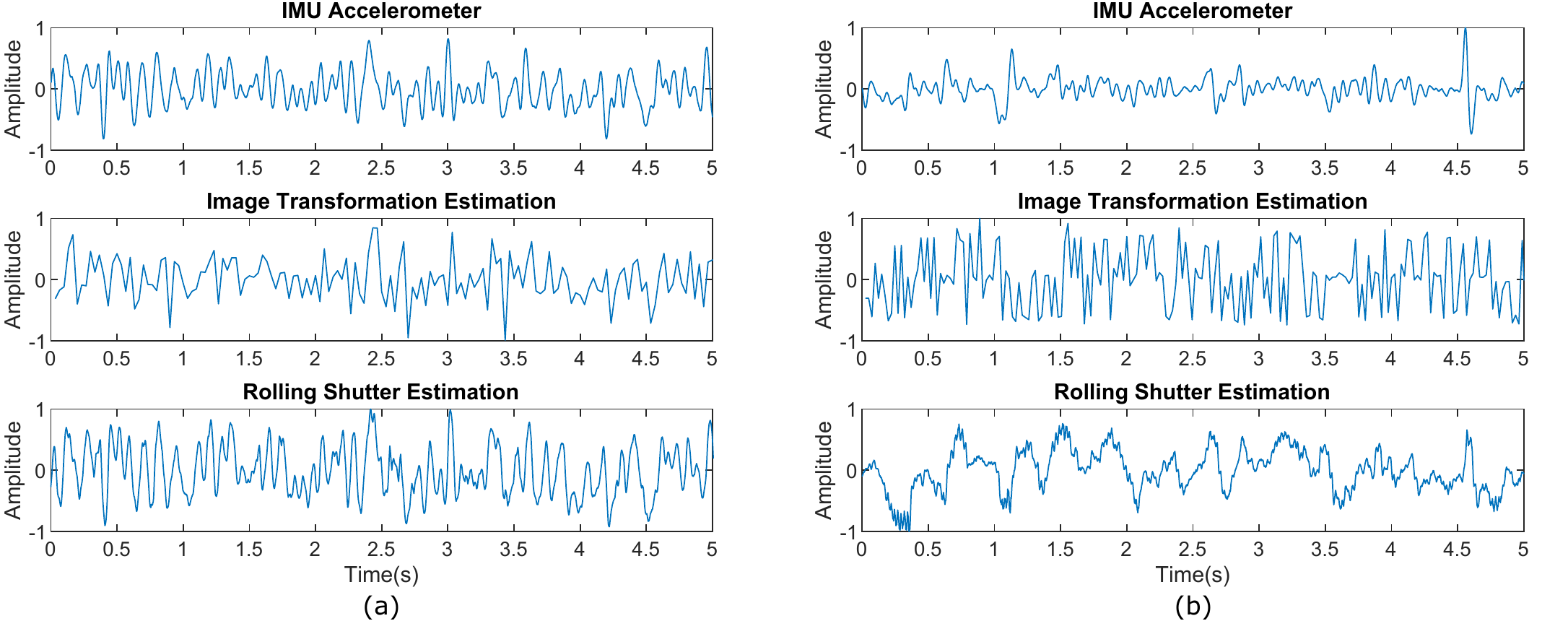}
	\vspace{-.1in}
	\caption{Measurements of physical IMU accelerometer (408 Hz) and virtual IMU  synthesized with the ITE and RSE methods from videos (30 fps frame rate, 1080p resolution) in 5 seconds. Amplitudes are normalized to compared different measurement approaches. (a) Videos stabilization is off. (b) Videos stabilization  is turned on. Strategically disabling sensor side channel mitigation mechanisms boosts up virtual sensors' capability for measurand authentication.} 
	\label{fig:corr}
\end{figure*}

\subsubsection{Low-frequency Information Measurement} \label{sec:decode_low}
The frame transformation perspective enables measurements of low-frequency components. It perceives the difference between two frames as the result of a single motion vector composed of single-axis motions (Figure~\ref{fig:trans}) within the period of one frame. The camera imaging process thus becomes the sampling process of the measurable motion signals with a sample rate that is the same as the video frame rate, e.g., 30 Hz in case of 30 fps videos. Theoretically, all image registration methods are applicable to extract inter-frame variations. We discuss one possible construction. 

\textbf{Image Transformation Estimation (ITE).} A straightforward way of extracting the frame differences is registering the frames with respect to a reference frame by estimating the 2D image transformations needed to warp the reference frame to the other frames as has been explored in~\cite{simon2013pin}. Each 2D transformation estimation generates a 3-by-3 transformation matrix. By concatenating each entry of different transformation matrices chronologically, it produces 9 vectors that represent the output of $\Tilde{f}_{\{v_i\}}$. Diverse algorithmic implementations of this method are possible.  This works uses an image registration implementation based on phase correlation~\cite{reddy1996fft}.

\subsubsection{High-frequency Information Measurement} \label{sec:decode_high}
The rolling shutter perspective allows for the extraction of intra-frame high-frequency variations. It perceives the difference between two frames as the result of multiple sequential motion vectors. The number of  motion vectors is the same as the number of rows of the camera imaging sensor as each row is exposed and sampled sequentially. The effective sample rate is thus the row-scanning rate of the rolling shutter, which is higher than 30 kHz for most commercial cameras. Nevertheless, not all signals within its Nyquist frequency can be recovered, as the non-zero exposure time causes motion blurs and attenuate the higher-frequency signals~\cite{davis2014visual}. Similarly, a possible construction is introduced below. 

\textbf{Rolling Shutter Estimation (RSE).}  Methods of rolling shutter estimation still compares different frames, but performs such comparison on the even smaller granularity level of rows or individual pixels. Then, the methods concatenate the values generated by the comparison first across different rows of a single frame, and then across different frames to form the motion signal vectors. With the proposed methodology, this work converts rolling shutter estimation into a pixel-level image registration problem. Algorithms capable of pixel-level registration often generate displacement fields, i.e., matrices of the same size as the registered images, on the X and Y directions. The produced matrices are apparently high-dimension and difficult to process. We can then group the matrices column-wise and  average the columns in each group  to produce easily understandable signals. This work uses a diffeomorphic image registration method~\cite{vercauteren2009diffeomorphic} to implement RSE.

\subsubsection{Demonstration}
Figure~\ref{fig:corr} shows the motion signals measured by a physical IMU (408 Hz sample rate) and virtual sensors using ITE and RSE methods.  A Google Pixel 2 smartphone held by a person recorded the physical IMU readings and camera videos simultaneously, where the postural hand tremor of the person caused the camera motion.  The ITE and RSE methods have sample rates of 30 Hz and 34 kHz respectively. The figure only displays a single vector of the physical and virtual sensor measurements respectively that represents the horizontal motion to simplify the visualization.   

Figure~\ref{fig:corr} (a) and (b) shows the measured signals with the video stabilization functionality being off and on respectively. When video stabilization is off, the virtual sensor outputs of both the ITE and RSE method show strong correlation with the physical IMU measurements. It is also clear that a 30 Hz sample rate is not sufficient to capture all the motion, as the ITE method's signal shows larger distortions than that of the RSE method. When video stabilization is turned on, the camera motion signals deviate more from the IMU readings as expected. Although the signal of RSE method still shows observable correlation with the IMU signal, ITE produces seemingly uncorrelated signals.

\subsection{Experiment}
We conduct preliminary tests with 4 people and a Google Pixel 2 smartphone. The 4 participants are all healthy males with similar ages, heights, and weights. As a proof-of-concept instead of an actual system product, we regard facial recognition and  tremor recognition as two decoupled problems and test them separately. The tremor recognition mechanism can be regarded as an additional layer of protection besides the existing facial recognition system. We investigate the impact of disabling and enabling video stabilization in both of the two tests.  

The objective of testing tremor recognition  is  to verify the effectiveness of the synthesized IMUs. To that end, we also recorded the physical IMU readings for comparison. The objective of testing facial recognition is two-fold. First, it is important to inspect if the postural hand tremor of different people can already make a difference in the original facial recognition systems without  synthesis of virtual sensors. This verifies the necessity of constructing dedicated virtual IMUs. Second, since turning off video stabilization may lead to better virtual sensor performance, it is also necessary to inspect if it would degrade the performance of facial recognition given that the videos are more shaky due to unmitigated camera motion.

\subsubsection{Data Collection.}
The 4 participants act as the legitimate user in turn and the remaining 3 participants act as the imposters. During the legitimate user sessions, each legitimate user holds the phone and records his own face for 30 times. We refer to these videos as legitimate videos.  During the spoofing attack sessions, each of the 3 imposters holds the phone but records the face of the legitimate user standing beside the imposter for 6 times to mimic a perfect silicone mask as assumed in Section~\ref{sec:case_primer}. We refer to these videos as imposter videos. Each video recording is about 6s in length and the physical IMU readings are recorded simultaneously.  The procedure is carried out first with video stabilization disabled. At the end, each participant recorded 48 videos when he held the phone with 30 of them being legitimate videos and the other 18 being imposter videos. We then repeat the procedure with video stabilization enabled. The total 384 videos (192 videos each set) are used for testing facial recognition and tremor recognition. 



\subsubsection{Test Procedure \& Result} We generalize the authentication problem as an identification problem and use classification models to measure the effectiveness of the two authentication schemes against the spoofing attack. 

\textbf{Facial recognition Procedure.}  We tested  MobileFaceNets~\cite{chen2018mobilefacenets} as the classification model which is a widely used facial recognition model designed for mobile platforms. 80\%  of each person's legitimate videos  are used to enroll their faces. The remaining 20\% of legitimate videos together with all imposter videos that contain faces of the legitimate users are used as the authentication test data. 

\textbf{Facial recognition Result.} Both the legitimate users and imposters' videos authenticated with 100\% success rate no matter the video stabilization was enabled or disabled. As expected, the results suggest that existing face authentication systems are mostly likely not designed to utilize camera  motion side channel information. Mapping it to Equation~\ref{eq:error_u}, it suggests $\mathbb{E}[\mathcal{A}_u(m) | H_0] \rightarrow 1$ and $\mathbb{E}[1-\mathcal{A}_u(m) | H_1] \rightarrow 0$ for the system under this specific spoofing attack. The results also show that disabling video stabilization to allow for more capable virtual IMUs did not affect the performance of the original facial recognition system. 

\textbf{Tremor Recognition Procedure.} For each video, we  generate virtual IMU measurements using both the ITE and RSE methods. We extract common time-domain and frequency-domain features as the ones used in \cite{miu2016person, bolton2021touchtone}. \rev{As a simple proof-of-concept, we did not use sophisticated machine learning models but directly utilized Matlab's implementation of  support vector machine (SVM) with a quadratic kernel and the default hyper-parameters~\cite{svm}. 5-fold cross validation was performed in the training phase along with a one-vs-one multi-class classification method.} Similar to facial recognition, \rev{for each legitimate user} we use 80\% of the legitimate videos \rev{(24 videos)} in the \rev{training}  phase and the remaining 20\% \rev{legitimate videos (6 videos)} together with all imposter videos \rev{(18 videos, 6 from each of the three imposters)} as authentication test data. We then calculate the true positive and  true negative rates on the test set. To provide comparisons, we repeat the same procedure also for the physical IMU data.

\textbf{Tremor Recognition Result.} Table~\ref{tab:tremor} shows the results of tremor recognition. Virtual IMU using RSE had performance approaching that of the physical IMU. It suggests that under this specific spoofing attack, $\mathbb{E}[\mathcal{A}_m([m_{_{int}}, m_{v_1}]) | H_0] \rightarrow 0.125$ and $\mathbb{E}[1-\mathcal{A}_m([m_{_{int}}, m_{v_1}]) | H_1]  \rightarrow 0.083$  if using an AND logic to combine facial and tremor recognition decisions. This results in $E_m - E_u \rightarrow -0.875c_1 + 0.083c_2$, which is highly likely to be smaller than 0. It is also clear that disabling video stabilization improves the performance of virtual IMUs. 

\color{revcol}

\subsubsection{Summary \& Implication}\label{sec:summary}

Our preliminary tests indicate a high probability that integrating user postural hand tremor information from camera motion side channels will help existing facial recognition systems defend against visual spoofing attacks. Test results show MobileFaceNets could recognize legitimate users with 100\% accuracy but could not detect (with 0\% accuracy) a powerful silicone mask spoofing attack that almost perfectly replicates visual features of users. This behavior is not a design defect of existing facial recognition systems, but an anticipated outcome of only using visual information during an authentication process. On the other hand, virtual IMUs synthesized from camera motion channel were able to detect such a visual spoofing attack with over 87.5\% accuracy at a cost of reducing true positive rate to 91.7\%. The simplest approach of integrating virtual sensor into existing facial recognition systems is to have a standalone tremor recognition module that processes camera motion information in the videos, and have the system declare a legitimate user only when both this tremor recognition module and the original facial recognition module declare it simultaneously. In this way, the overall system's security performance increases in the face of facial spoofing attacks even with a lower true positive rate. This result also suggests \textit{when a physical sensor system has poor performance on a security task, it is easy to produce an obvious marginal benefit on the system's performance by integrating sensor side channel information.} Of course, a more sophisticated decision system can tune its weights on the facial and tremor recognition modules to strike a  better balance between usability and security. 

Beyond camera motion side channels, our tests also provide one viable data point for the general concept of utilizing sensor side channels and reveal some common problems it faces. For example, we expect the same problem of usability-security trade-off in using virtual sensors synthesized from sensor side channels alongside the original physical sensors. Essentially, physical sensors and synthesized virtual sensors provide two streams of information, each one of which is more reliable in one task but also unreliable in another task. The design trade-off appears when the overall system needs to complete both tasks to achieve its functionality. 

\subsubsection{Limitation \& Future Work.} With the goal of showing a proof-of-concept example, our experiment  provides empirical statistical evidence for the benefit of utilizing camera motion side channels only based on a very limited data distribution. The limitations of tested data lie in the following 4 main dimensions.

First, the 4 young male participants  may not provide a high enough degree of demographic diversity, especially for evaluating postural hand tremors which are highly dependent on age, gender, and health conditions~\cite{hubble1997clinical}. While we based our choice of the 4 participants on the hypothesis that more similar participants produce less distinct tremor patterns and thus help us estimate a lower bound of tremor recognition performance, we believe studying more diverse groups of people will generate new insights into recognition performance variability and possible strategies of recognition algorithm design.  

Second, we collected 30 samples of legitimate-user videos and 18 spoofing attack videos for each legitimate user’s authentication session within a single day. We find this initial set of samples provided evidence to suggest the potential of utilizing hand tremor information from camera side channels to enhance existing facial recognition system’s security. It is possible that tremor patterns can change with time. Although previous research shows hand tremor remains stable after 78 days~\cite{dun2019replicability}, a longer duration needs to be investigated in future complete. The recognition system may need to periodically update its database if tremor pattern is found to vary over time. 

Third, we emulated perfect silicone masks by using the real faces of legitimate users. This only provides an estimate of the upper bound of the overall recognition system's performance improvement when tremor recognition is used. Specifically, the benefit of including tremor recognition may get lower when a worse-quality silicone mask is used because the damage the attack can do to the original unimodal authentication system is lower while tremor recognition still causes a decrease in the true positive rate. As a result, we suggest future works  test different qualities of silicone masks on popular facial recognition systems to better assess the benefit of including virtual IMUs for tremor recognition. 

\camready{Fourth, the decoupling of facial recognition and tremor recognition problems in this proof-of-concept case study prevents us from utilizing the temporal correlation between the facial and camera motion signals and investigating the impact of the correlation information. Intuitively, systems that inspect such temporal correlation information require spoofing attackers to further achieve synchronization between the physical and virtual sensors' data streams and thus provide additional protection. We envision real-world products building upon the virtual sensors authentication concept to utilize deep-learning approaches for processing temporally-correlated physical and virtual sensors' information.} 

\color{black}





\begin{table}
\centering
\caption{Test Accuracy of Tremor Recognition}\label{tab:tremor}  
\vspace{-.15in}
\begin{tabular}{|c|c|c|c|c|c|c|} 
\hline
\multirow{2}{*}{} & \multicolumn{2}{c|}{Physical IMU} & \multicolumn{2}{c|}{Virtual ITE} & \multicolumn{2}{c|}{Virtual RSE} \\ 
\cline{2-7}
 & TPR & TNR & TPR & TNR& TPR & TNR  \\ 
\hline
Stab. OFF & 95.8\% & 94.4\% & 62.5\% & 65.3\% & 91.7\% & 87.5\% \\ 
\hline
Stab. ON & 95.8\% & 93.1\% & 45.8\% & 41.7\% & 70.8\% & 72.2\% \\
\hline
\end{tabular}

\vspace{.05in}

\raggedright \setstretch{0.5}  \small
\rev{TPR (true positive rate) and TNR (true negative rate) are the percentages of correctly recognizing a legitimate user and a perfect silicone mask spoofing attack respectively. In comparison,  MobileFaceNets had  TPR=100\% and TNR=0\% in our test.}   \label{tab:tremor}  

\end{table}

\section{Discussion}

Below we discuss the major areas of possible future work and interesting research questions.

\textbf{Sensor Side Channel Models.} \rev{To support future applications of sensor side channels, we believe  more concrete and computable mathematical  models than the framework proposed in Section~\ref{sec:prob} are needed as the current framework relies on abstract concepts instead of rigorous mathematical derivations.  We envision future models to have the following features. First, they need to enable exact definitions and determination of different types of sensor side channels by providing the algorithms for calculating signal correlations and threshold values. Second, they need to provide quantitative metrics for measuring the usability-security trade-off mentioned in Section~\ref{sec:summary}. 
}
\camready{Third, they need to delineate  mechanisms for measuring the available signal quality and bandwidth of side channel measurement components.}

\textbf{Security for Sensor Side Channel Authentication.} Technically, inseparable sensor side channels also provide the information needed for measurand authentication. We advocate the use of separable and controllable sensor side channels because they are protected from adversaries that exploit unmitigated side channels (Section~\ref{sec:bg_attack}).  Nevertheless, risks of malicious exploitation still exist within authentication time. It is thus necessary for future works to consider how to ensure that side channels benefit the
defender, but not adversaries that attempt eavesdropping and transduction attacks, during authentication. 

We believe an access control and permission system that is similar to existing systems managing physical sensors on mobile platforms (e.g., Android) can be employed to prevent eavesdropping attacks. Virtual sensor entries can potentially be created and integrated into existing permission systems so that knowledge and  methodology of solving physical sensors' problems can also benefit virtual sensors. Transduction attacks, on the other hand, are harder to address. In the context of sensor side channel based measurand authentication, transduction attacks can be generalized as authentication spoofing that tries to modify perceived characteristics of the byproducts, measurers, and environments. As a result, existing methodologies of spoofing detection may be applied. In summary, we believe there are opportunities to address the problems of virtual sensors by reflecting on existing methodology for physical sensors.  

\color{crcol}

\textbf{Side Channels vs. Legitimate Channels.} We believe there will be an interesting phenomenon that sensor side channels  are turned into legitimate communication channels when  active controls and dedicated APIs  are developed to support as well as  regulate the use of sensor side channels in the future.  After all, the key difference between side channels and legitimate channels is whether the channels are designed, intended, and allowed by the system's security policy or not. When such side channels are regarded as legitimate channels, however, new side-channel information may again be discovered to be embedded in such ``legitimate'' information as hardware and computation technologies keep advancing and extending the boundary of recoverable physical signals. We thus believe it is necessary for researchers to take a development perspective and periodically examine the security implications of sensor side channels.

\textbf{Fewer Sensors via Sensor Repurposing.}  In a broader context, we believe the technique of synthesizing virtual sensors from sensor side channels aligns with the general idea of repurposing sensors for different sensing tasks. Essentially, we are trying to shift sensor hardware functionalities to the software space by understanding the transformation between different forms of signal energy and carrying out additional model-based computations. In contrast to the current trend of deploying more and more sensors in the Internet of Things era, we cannot help thinking if such sensor repurposing ideas would allow us to reduce the number of physical sensors and achieve more abstract and manageable sensor peripheral systems that are subjected to smaller attack surfaces. 

Besides reducing the number of physical sensors, the technique could also be applied to enhance existing systems that require new functionalities but have harsh environmental conditions where a hardware update is challenging. This idea is revealed in the example of NASA’s Voyager 1  spacecraft which needed to measure plasma density in order to determine its location relative to the heliosphere. Voyager 1's plasma spectrometer stopped working in 1980, making a direct plasma density measurement impossible. However, the operation team learned that our sun sometimes emits shock waves that can  cause the plasma surrounding the spacecraft to oscillate. The team then measured the oscillation  using  Voyager 1’s onboard plasma wave sensing system as a proxy  of the plasma density~\cite{gurnett2013situ}, essentially synthesizing a virtual plasma density sensor by understanding the energy transformations. 

\color{black}

\section{Conclusion}

This paper argued that analog sensor side channels can benefit defenders by providing an opportunity to authenticate the sensor measurands. Future sensor designs can consider actively controlling sensor side channels after finding ways to mitigate these channels, instead of simply eliminating sensor side channels. We first introduced a framework for defining and characterizing sensor side channels, and then formulated the problem of measurand authentication using virtual sensors synthesized from sensor side channels. We also introduced three specific ways of applying the model of measurand authentication by verifying signal byproducts, sensor measurers, and sensor environments respectively, and provided examples of each case. 

\camready{Synthesizing virtual sensors from the side channels of physical sensors formulates a  mechanism for repurposing existing sensor hardware to harvest extra modalities of information. We believe the applications of this mechanism can potentially span a much larger scope than authentication. Going forward, we envision that virtual sensor synthesis could develop into a new research area that actively interacts with the existing research areas of digital forensics, sensor fusion, multimodal deep learning and perception, etc. The fundamental research question we will need to explore is how to model the transformations between the energies of different information modalities.} 

\bibliographystyle{ACM-Reference-Format}
\bibliography{refs}

\end{document}